\documentclass[useAMS,usenatbib]{mn2e}
\usepackage[pdftex]{graphicx}
\usepackage{mesjournaux}
\newcommand{\be}{\begin{equation}}
\newcommand{\ee}{\end{equation}}

\title[Fan-shaped winds from Keplerian accretion discs]{On fan-shaped cold MHD winds from Keplerian accretion discs}  
\author[J. Ferreira and F. Casse]{J. Ferreira$^{1}$\thanks{Jonathan.Ferreira@obs.ujf-grenoble.fr} and F. Casse$^{2}$\thanks{fcasse@apc.univ-paris7.fr}\\
$^{1}$ UJF-Grenoble 1/CNRS-INSU, Institut de Plan\'etologie et d'Astrophysique de Grenoble (IPAG) UMR 5274, Grenoble, F-38041, France\\
$^{2}$ Laboratoire AstroParticule \& Cosmologie (APC), Universit\'e Paris Diderot, CNRS/IN2P3, CEA/Irfu, Observatoire de Paris,\\ Sorbonne Paris Cit\'e  -   \ 10, rue Alice Domon et L\'eonie Duquet, F-75205 Paris Cedex 13, France}
\begin{document}

\maketitle

\begin{abstract}

We investigate under which conditions cold, fan-shaped winds can be steadily launched from thin (Keplerian) accretion discs. Such winds are magneto-centrifugal winds launched from a thin annulus in the disc, along open magnetic field lines that fan out above the disc. In principle, such winds could be found in two situations: (1) at the interface between an inner Jet Emitting Disc, which is itself powering magneto-centrifugally driven winds, and an outer standard accretion disc; (2) at the interface between an inner closed stellar magnetosphere and the outer standard accretion disc. We refer to Terminal or T-winds to the former kind and to Magnetospheric or M-winds to the latter.     

The full set of resistive and viscous steady state MHD equations are analyzed for the disc (the annulus), which allow us to derive general expressions valid for both configurations. We find that, under the framework of our analysis, the only source of energy able to power any kind of fan-shaped winds is the viscous transport of rotational energy coming below the inner radii. Using standard local $\alpha$ prescriptions for the anomalous (turbulent) transport of angular momentum and magnetic fields in the disc, we derive the strength of the transport coefficients that are needed to steadily sustain the global configuration.  It turns out that, in order for these winds to be dynamically relevant and explain observed jets, the disc coefficients must be far much larger than values expected from current knowledge of turbulence occurring inside proto-stellar discs.  

Either the current view on MHD turbulence must be deeply reconsidered or steady-state fan-shaped winds are never realized in Nature. The latter hypothesis seems to be consistent with current numerical simulations. 

\end{abstract}


%

\section{Introduction}

\subsection{Mass loss from young stars}

Young Stellar Objects (hereafter YSOs) show evidences of fast (several hundred km/s) and collimated jets that carry a sizable fraction of the released accretion power (e.g. \citealt{ball07,ray07}). It has been soon recognized that large scale magnetic fields play a major role in driving these jets. Basically, this is the only process that has been proven to be able to simultaneously provide a very efficient acceleration along with a self-confinement of the ejected plasma (see \citealt{cabr07} for a review). These large scale magnetic fields can be anchored onto the star (giving birth to a "stellar wind", e.g.  \citealt{hart82a, saut94, matt05a}), the accretion disc ("disc wind", e.g. \citealt{blan82,pudr86, ward93, ferr93a}) or linked to both ("X-winds", \citealt{shu94a,cai08} and "Reconnection X-winds", \citealt{ferr00}). 

Each model corresponds actually to a different scenario for the origin of the magnetic field. On one extreme, disc winds rely on the existence of a large scale magnetic field threading a significant portion of the disc (up to $\sim 100$ AU for instance in earlier works). On the other extreme, stellar winds and X-winds assume the presence of such large scale fields only on the star. Since direct observations of magnetic fields in discs are highly difficult (see for instance \citealt{dona05}), some conjectures must be made. However, these wind models are described by the same set of magnetohydrodynamic (hereafter MHD) equations and can be easily confronted to optical and near-IR observations of YSO jets \citep{ferr06b}. For instance, if the transverse velocity shifts observed in some of these jets are indeed due to jet rotation, then stellar winds and X-winds (in fact any wind coming from the innermost regions) are ruled out as the main jet component: they simply do not carry enough angular momentum to explain observations \citep{ande03,ferr06b}. But on the other hand, these optical observations imply also that the last relevant field lines in disc winds cannot be anchored farther than a few AU (for the atomic component: molecular jets could be anchored up to ten AU, \citealt{pano12}). Thus, whether YSO jets are rotating or not, observational evidences suggest that jets are arising from regions smaller than or of the order of several AU, while discs could be as large as several hundreds of AU.  

\begin{figure*}
\centering
\[ \begin{array}{cc} 
 \includegraphics[width=0.5\textwidth]{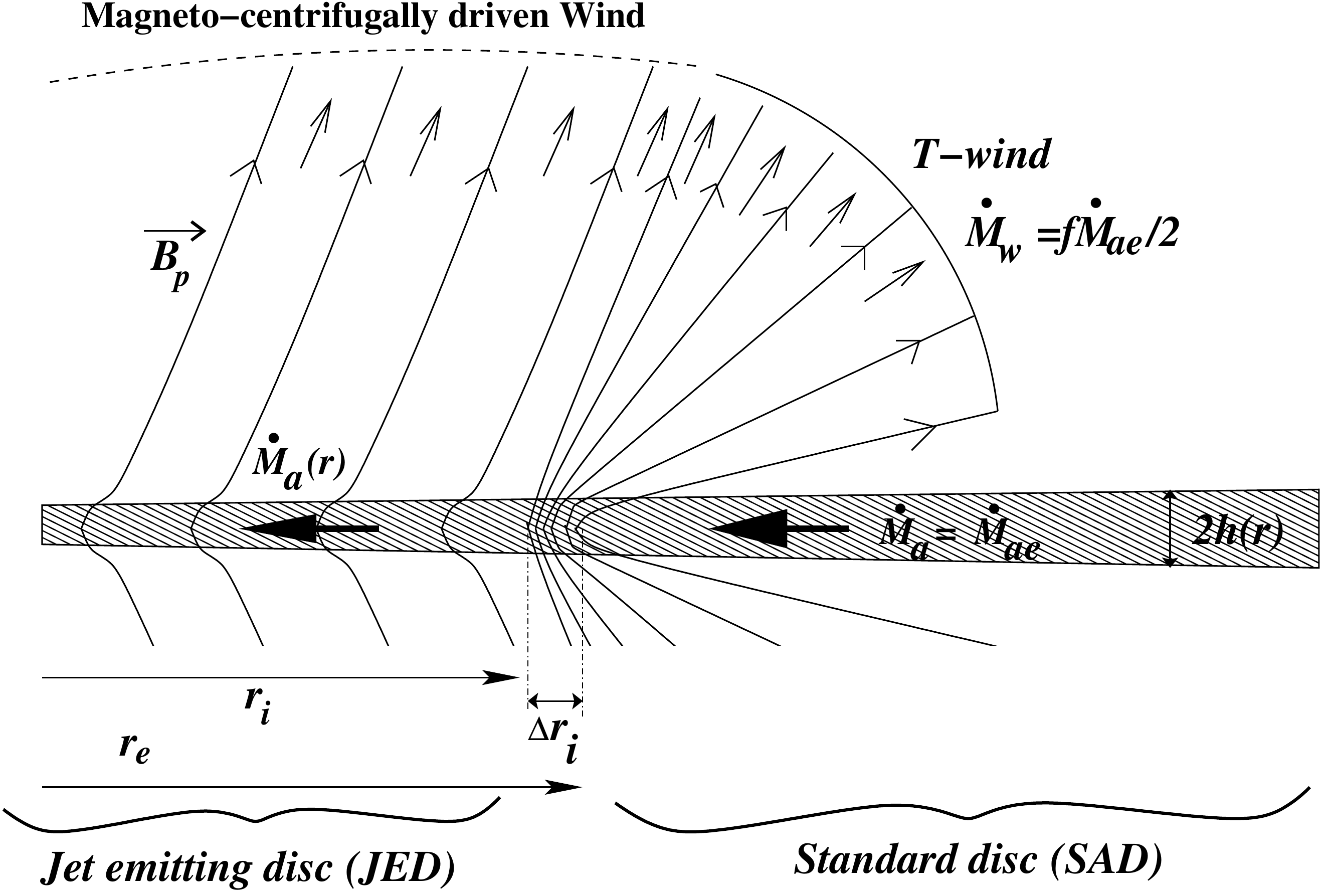} & 
  \includegraphics[width=0.5\textwidth]{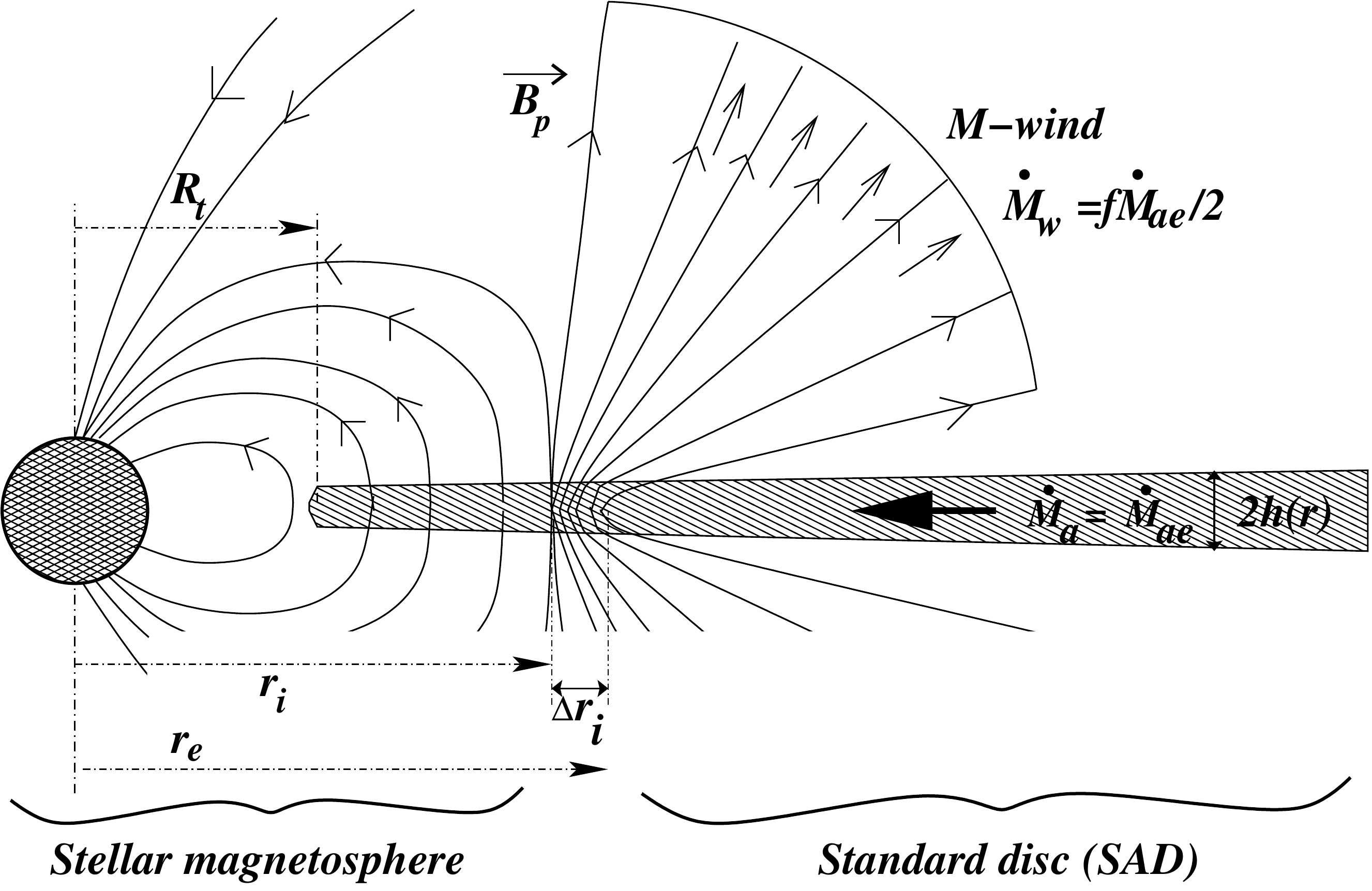}
  \end{array}  \]
 \caption{The two configurations of fan-shaped winds. {\bf Left}: Terminal or T-winds launched from $r_i= r_J$, the outer radius of a JED, to $r_e= r_i + \Delta r_i$ the inner radius of a standard accretion disc (SAD). The T-wind leans against the inner magneto-centrifugally driven wind and fans out above the standard disc.
 {\bf Right}: Magnetospheric or M-wind at the interface between an outer standard disc and the inner magnetosphere of a central star, with $r_i \simeq r_{bf}$, where $r_{bf}$ is the base of the funnel flow \citep{bess08}. In this configuration, three cases can be realized according to the strength of the stellar magnetic dipole as measured by the ratio of the truncation radius $R_t < r_i$ to the co-rotation radius $R_{co}$: M-winds with $r_i < R_{co}$, M-winds with $r_i > R_{co}$ and X-winds, which are M-winds with $r_i \simeq R_{co}$.}
\label{F1}
\end{figure*}

\subsection{What are fan-shaped winds?}

If observed YSO jets are indeed mainly fed by disc winds, then the mass, angular momentum and energy extraction due to the winds have a profound impact on the dynamics (and possibly observational appearance) of the underlying Jet Emitting Disc (hereafter JED, \citealt{comb08, comb10}). It has been analytically shown \citep{ferr95, ferr97,cass00a} and numerically confirmed \citep{tzef09} that the large scale magnetic field must be close to equipartition in order to magnetically launch jets in a steady fashion. A natural transition from an inner JED to an outer standard accretion disc will therefore occur if, beyond a radius $r_J$, the local disc magnetization suddenly decreases. If this drop is done over a thin radial extent, namely $\Delta r \sim h \ll r$ where $h$ is the local disc vertical scale height, then the poloidal magnetic field tends to take a fan-like geometry above the disc (see Fig.~\ref{F1}a). This region of highly bent, open magnetic field lines could also be the locus of some ejection, allowing thereby to act as a transition region between the outer non ejecting disc to the inner JED. These MHD winds will be hereafter referred to as "Terminal-winds" or T-winds. 

A different sort of fan-shaped MHD wind could also be found at the interface between an outer standard accretion disc (with no or negligible large scale magnetic field) and the stellar magnetosphere. For simplicity, let us consider only the case of a stellar magnetic moment aligned with the rotation axis of the star and the disc. The disc is truncated at a radius $R_t$ where the stellar magnetic field is near equipartition with the disc thermal pressure \citep{roma02,bess08, zann09, roma11,roma12}. Such an interaction leads to the opening of stellar field lines so that some magnetic flux diffuses in the disc. Because of the sharp decrease of the (initially) stellar dipole field, the zone where this field is dynamically important is very thin, a typical length being $\Delta r$ of order of a few scale heights. Here again, such a geometry appears to be a favorable place for magnetically launching winds (see Fig.~\ref{F1}b). These  "Magnetospheric-winds" or M-winds are anchored in a radial zone that could act as a buffer zone between the outer non-ejecting disc and the inner magnetospheric accretion curtains.

In the case of M-winds, three different situations can take place depending on the location of the disc truncation radius $R_t$ with respect to the co-rotation radius  $R_{co}$ (defined as the radius where the Keplerian $\Omega_K$ and stellar $\Omega_*$ angular velocities are equal). When $R_t < R_{co}$ accretion columns will be formed whereas a "propeller regime" is settled when  $R_t > R_{co}$ (\citealt{roma03a,roma05, usty06} and references therein). The limiting case where $R_t \simeq R_{co}$ is known in the literature as "X-winds" and its axisymmetric, steady state ideal MHD wind structure has been explored in several papers \citep{shu94a, shu94b, naji94, shu95, shan98, shan02, cai08}. The computed 2D flow structure is a fan-shaped cold wind expanding from a point source with an initially super slow-magnetosonic speed that is matched to an asymptotic jet cylindrical solution. In these wind calculations the underlying disc is a mere boundary condition and the impact of the winds on the underlying disc dynamics has not been investigated yet. 

The goal of this paper is to analyze at which conditions fan-shaped winds, namely T-winds and M-winds, can be launched from a near Keplerian accretion disc. By fan-shaped wind, we mean {\em any} steady-state axisymmetric MHD wind that originates from a region of extent $\Delta r \sim h$ in a Keplerian accretion disc and with the tendency to fill in all space (unless confined by an outer pressure). The questions we plan to address are the following: how much mass and power can be carried away by these winds?  Is a steady state possible? What are the constraints imposed on the underlying MHD turbulence in the disc? 

In Section~2, we write down and analyze all steady-state MHD equations describing the disc and its winds. Note that the disc is actually a thin annulus, from $r_i$ to $r_e= r_i + \Delta r_i$ with  $\Delta r_i \sim h$. Our results will be general, valid for any kind of fan-shaped wind. Then, in Section~3 we apply them to each specific case, namely T-winds and M-winds, and address the questions raised above. The caveats and consequences of our analysis will be discussed in Section~4.

\section{General aspects of discs driving cold, fan-shaped winds}

In this Section, we write down all equations and put them into a form that is general and valid for the two envisioned fan-shaped wind scenarios, T-winds and M-winds. Next Sections will be devoted to each of these scenarios by applying our general results to each specific case.    

\subsection{Governing MHD equations for the disc}
To describe the disc, the following set of non-relativitistic, axisymmetric, steady-state single fluid MHD equations  must be analyzed
\begin{eqnarray}
\label{eq:mass} & & \nabla \cdot \rho {\bf u}   =  0 \\ 
\label{eq:mom} & & \rho {\bf u} \cdot\nabla {\bf u} =  - \rho \nabla \Phi_G\; - \;\nabla P 
\;+ \; {\bf J} \times {\bf B}\; + \; \nabla\cdot \mathbf T \\ 
\label{eq:diff}& & \eta_m J_{\phi}  =  {\bf u}_p \times {\bf B}_p \\
\label{eq:ind1} & & \nabla \cdot ( {\nu'_m \over r^2} \nabla r B_{\phi})  =   \nabla
\cdot{1\over r} (B_{\phi}{\bf u}_p - {\bf B}_p\Omega r )
\end{eqnarray}
\noindent where ${\bf J}  =  \nabla \times {\bf B}/\mu_o$ is the electric current density, $\Phi_G= - GM/(r^2 + z^2)^{1/2}$ is the gravitational potential of the central object and other quantities have their usual meaning. Here,  $\mathbf T$ is the turbulent stress tensor resulting from a self-sustained MHD turbulence within the magnetized disc. Following \citet{shak73}, the only relevant component used here is $T_{r\phi}=\displaystyle\rho\nu_vr \left(\frac{\partial\Omega}{\partial r}\right)$. Our mean field approach assumes that the disc turbulence provides anomalous transport coefficients such as a turbulent viscosity $\nu_v$ and magnetic diffusivity $\nu_m$ ($\eta_m=\mu_o \nu_m$ is the resistivity). This diffusivity, appearing in the toroidal component of Ohm's law (\ref{eq:diff}), allows the diffusion of mass across the poloidal magnetic field ${\bf B}_p$ and thereby the possibility of a steady state. A diffusion process must also be present with respect to the toroidal motion, otherwise the field would be overly sheared with a tremendous toroidal field component $B_{\phi}$. In the induction equation (\ref{eq:ind1}), some turbulent magnetic diffusivity $\nu'_m$ has to be introduced, so that the dimensionless number 
\be
\chi_m = \frac{\nu_m}{\nu'_m}
\ee
measures a possible anisotropy of the turbulent diffusion processes within the disc \citep{cass00a}. The amplitude of the turbulent transport coefficients are measured using $\alpha$ coefficients, namely $\alpha_v =\nu_v /\Omega_k h$, $\alpha_m = \nu_m/\Omega_K h$ and $\alpha'_m=\nu'_m/\Omega_K h$, so that the effective magnetic Prandtl number writes
\be
{\cal P}_m = \frac{\nu_v}{\nu_m} = \frac{\alpha_v}{\alpha_m}
\ee
These $\alpha$ coefficients are expected to decrease vertically on a disc scale height, so that the flow (the wind) becomes inviscid and described by ideal MHD equations. It is noteworthy that these $\alpha$ coefficients can also vary radially.
Indeed, their magnitude depend on, e.g. the field strength and the degree of ionization, both effects that vary throughout the disc. For instance, one might have the viscosity parameter $\alpha_v $ going from a very low value at the disc outer edge (cold disc and low magnetic fields) to a much larger value (possibly reaching unity) at the disc inner edge. On the other hand, while the degree of ionization has a critical impact on the development of MHD turbulence (see eg. \citealt{lesu07,bai11} and references therein), its radial dependency is unknown and depends on the assumptions made about the disc density structure (see for instance \citealt{from02,comb10}). To circumvent these uncertainties, we will hereafter assume that ${\cal P}_m $ undergoes no significant change within the portion of the disc considered here. This is clearly a limitation of our model. 

Finally, in order to close the above set of equations, we will assume a perfect gas law in the disc, namely $P= \rho C_s^2$, with a vertical isothermal approximation. This latter assumption is not critical for our purpose as we deal with cold ejection from near Keplerian accretion discs. 

The dynamical importance of the large scale magnetic field is measured at the equatorial plane by the disc magnetization 
\be
\mu = \left . \frac{B_z^2}{\mu_o P_{tot}} \right |_{z=0}
\label{Eq:defmu}
\ee 
where $P_{tot}$ is the total (gas + radiation) pressure in the disc \citep{ferr95}. In a disc with a negligible radiation pressure as considered here, $\mu = 2/\beta$ where $\beta$ is the usual beta plasma parameter. Maintaining a disc with a Keplerian profile despite the presence of large scale magnetic fields puts an upper limit on the magnetization: $\mu$ cannot be larger than unity \citep{ferr95,ferr97,shu08}. Under these circumstances, the magnetic pinching force is at most comparable to the gravitational component and the disc aspect ratio can still be approximated by
\be
\varepsilon = \frac{h}{r} \sim \left .  \sqrt{ \frac{P}{\rho \Omega_K^2 r^2} }  \right |_{z=0} \simeq \frac{C_s}{\Omega_K r}  \ll 1
\label{eq:h}
\ee
namely $C_s \sim \Omega_K h$, as in a standard accretion disc. The actual scale height of the disc will be smaller than, but of the order of $h$, which is enough for our purpose here (see Appendix \ref{AppB} for more details).

\subsection{Cold winds and magnetic field configuration}
Fan-shaped winds are assumed to be cold, that is magnetically driven. This implies that the magnetic force must be the dominant one above the disc and that enthalpy can be neglected. A successful magnetic ejection then requires some conditions at the disc surface that can be translated into conditions on the magnetic field geometry, namely on the radial $B_r^+$ and toroidal $B_\phi^+$ magnetic field components at the disc surface. 

The energetic condition to launch cold jets from near Keplerian discs has been first derived by \citet{blan82}. It states that the poloidal magnetic field lines at the disc surface must be inclined by more than $30\deg$ with respect to the local vertical. This translates into
\be
p = \frac{B_r^+}{B_z}
\ee
larger than, but of the order of unity, where $B_z$ is assumed to remain of the order of the field at the midplane (see Appendix \ref{AppA}). One simple way to look at the ejection process is to project the magnetic force along a magnetic poloidal surface, which gives
\begin{eqnarray}
F_{\phi} & = & ({\bf J} \times {\bf B})\cdot {\bf e}_\phi = \frac{B_p}{2\pi r} \nabla_{\parallel} I \nonumber \\
F_{\parallel} & = & ({\bf J} \times {\bf B})\cdot {\bf e}_\parallel =   - \frac{B_{\phi}}{2\pi r}\nabla_{\parallel} I
\end{eqnarray}
where ${\bf B}_p= B_p {\bf e}_\parallel$, $\nabla_{\parallel}= {\bf e}_\parallel \cdot \nabla$  and $I = 2 \pi r B_{\phi}/\mu_o <0 $ is the total current flowing within this magnetic surface \citep{ferr97}. These expressions show that plasma acceleration along any given magnetic surface can be seen as a current leakage through this surface, providing simultaneously an azimuthal and a poloidal acceleration. In other words, while the magnetic field is both braking down ($F_{\phi} <0$) and vertically pinching ($F_{\parallel} \simeq F_z <0$) the disc, the sign of these two forces must change at the surface so that ejection becomes possible \citep{ferr93b, ferr95}.  This can be naturally achieved but imposes a constraint on the magnetic shear so that 
\be
q=  - \frac{B_\phi^+}{B_z}  \simeq \frac{1}{\alpha'_m}
\ee
This important condition is derived from the MHD induction equation (\ref{eq:ind1}) applied in a Keplerian disc where the dominant contribution is $B_r \partial \Omega/\partial r$ (because of the magnetic geometry considered here, see Appendix \ref{AppC} for details).

\subsection{The disc angular momentum transport}
\label{sec:mom}
The disc angular momentum equation ($\phi$-component of Eq.~\ref{eq:mom}) writes
\be
\frac{\rho {\bf u}_p}{r} \cdot \nabla \Omega r^2 = F_\phi +  \frac{1}{r^2} \frac{\partial}{\partial r} \rho \nu_v r^3 \frac{\partial \Omega}{\partial r} 
\ee
where $F_\phi=  J_z B_r- J_r B_z$. Integrating this equation over the disc thickness gives 
\be
\frac{\dot M_a \Omega_K}{4\pi r}  \simeq - 2 \frac{B_zB_\phi^+}{\mu_o} + \frac{3}{2 r^2} \frac{\partial}{\partial r}  \Sigma \nu_v \Omega_K r^2
\ee 
where both $\Omega \simeq \Omega_K$ and $h$ roughly constant over the extent $\Delta r_i$ have been assumed. Here, $\Sigma$ stands for the local disc surface density, defined as $\Sigma = \int_{-h}^{h}\rho dz$. Defining the  accretion sonic Mach number as $m_s= \dot M_a/2\pi \Sigma \Omega_K rh= u_o/C_s$, where $u_o$ is the average accretion velocity, one gets
\be
m_s =  m_s^{mag} + m_s^{visc} = 2 q \mu +  \zeta \alpha_v \varepsilon
\ee
where $m_s^{mag}$ and $m_s^{visc}$ are respectively the magnetic and viscous contributions and where
\be
\zeta =  3 \frac{d \ln (\Sigma \nu_v r^{1/2})}{d \ln r} = \frac{3}{2} - 3 \frac{d \ln {\cal R}_e}{d \ln r} + 3 \frac{d \ln \dot M_a}{d \ln r} 
\label{eq:zeta}
\ee
Here, the vertically averaged Reynolds number of the accreting flow writes ${\cal R}_e= r u_o/\nu_v= m_s/\alpha_v \varepsilon= \zeta( 1 + \Lambda)$, where $\Lambda= m_s^{mag}/m_s^{visc}$ is the ratio of the magnetic to the viscous torque. In a standard accretion disc (hereafter SAD), the magnetic torque is negligible and ${\cal R}_e= \zeta= 3/2$. In a JED, where the accretion rate varies as $\dot M_a \propto r^\xi$ \citep{ferr93a}, one gets $\zeta= 3/2 + 3 \xi$. Note also that, in a steady-state picture, the viscous torque is always braking down the disc plasma unless ${\cal R}_e$ increases with the radius. In usual accretion disc models, ${\cal R}_e$ is however a constant.

\subsection{Properties of cold fan-shaped winds}
\label{sec:mu}
Above the disc, turbulence is assumed to decay and ideal MHD applies. Steady-state, axisymmetric cold MHD jets can be described with 4 invariants: magnetic flux to mass flux ratio $\eta$, surface angular velocity $\Omega_B$, total specific angular momentum $L= \Omega_B r_A^2$ (where $r_A$ is the Alfv\'en cylindrical radius) and specific energy $E$. Our objective in this section is to relate these invariants to the disc physics, allowing henceforth to link cold fan-shaped wind asymptotic properties to disc properties.  

Mass and magnetic flux conservation provide
\be  
{\bf u}_p = \frac{\eta (\Phi)}{\mu_o\rho}{\bf B}_p
\label{eq:MHDI1}
\ee 
\noindent where $\eta(\Phi)$ is a constant along a magnetic surface defined by a constant $\Phi = \int 2 \pi r B_z dr$. It measures the ratio of the mass flux to the magnetic flux carried in by the wind, namely $\dot M_w =  \int 2 \pi r \rho u_z dr =  \int  \frac{\eta }{\mu_o}  d\Phi$, where the integration is made over the total magnetic flux $\Phi_X$ leaving the region of extent $\Delta r_i$. 

The induction equation (\ref{eq:ind1}) becomes in ideal MHD
\be 
\Omega_*(\Phi) = \Omega -\eta\frac{B_{\phi}}{\mu_o\rho r} \ ,
\label{eq:MHDI2}
\ee 
\noindent where $\Omega_*$ is the rotation rate of a magnetic surface, assumed here to be the Keplerian value at $r_i$. Because of plasma inertia, magnetic field and plasma have not the same rotation rate, which induces a toroidal field component.  Plasma velocity can be written ${\bf u} = (\eta/\mu_o\rho)  {\bf B} + \Omega_*r{\bf e}_{\phi}$ and is not parallel to the total magnetic field. In the reference frame rotating at $\Omega_*$ however, they do become parallel. 

The only explicit calculations of stationary fan-shaped MHD flows that can be found in the literature are those related to the X-wind \citep{naji94, shu95, cai08}. To ease comparison with these studies, we will use hereafter the same notations, with $R_X = r_i$. 
Defining fiducial quantities  $\dot M_X = \dot M_w$, $\rho_X=\dot M_X/2\pi r_X^3\Omega_X\ $, $V_X=\Omega_X R_X$,  $B_X=(\mu_o\Omega_X\dot M_X/2\pi R_X)^{1/2}$ and  $\Omega_X= (GM/R_X^3)^{1/2}$ accordingly to \citet{shu94a}, the dimensionless velocity and magnetic field are related by $ {\bf B} = \beta(\Phi)  \rho {\bf u}$ where
\be
\beta(\Phi) = \frac{\mu_o \rho_X V_X}{\eta B_X} = \kappa_{BP}^{-1} \frac{B_X}{B_{z,i}}
\ee 
In this expression, $\kappa_{BP}= \eta \Omega_{K,i} r_i/B_{z,i}$ is the \citet{blan82} mass loading parameter. Using $\dot M_w = \int d \dot M_w = \int d\Phi \rho_X V_X/\beta B_X$ one gets
\be
2 \pi B_X R_X^2 = \int_0^{\Phi_X} \frac{d\Phi}{\beta(\Phi)} \simeq \frac{\Phi_X}{\bar \beta}
\label{eq:Phi}
\ee
where $\bar \beta$ is some average value in the wind\footnote{If we define $\Phi= \Phi_X \Psi$ and take the scaling $\beta(\Psi)= \frac{2}{3}\bar \beta (1-\Psi)^{-1/3} $ used by \citet{cai08},  the rhs of Eq.(\ref{eq:Phi}) is multiplied by 9/8.}. 
 Since this magnetic flux is arising from a tiny disc region around a radius $r_i$, we can estimate it as $\Phi_X \simeq 2 \pi B_{z,i} r_i \Delta r_i$, where $B_{z,i}$ is the vertical magnetic field. Its value is then related to the fiducial field $B_X$ by $B_{z,i} \simeq  \bar \beta B_Xr_i/\Delta r_i$. This is the cornerstone of fan-shaped winds: a spherical dilution of the magnetic field, giving an average mass loading parameter $\bar \kappa_{BP}= \bar\beta^{-2}\Delta r_i/r_i$. Using the definition of the disc magnetization at $r_i$ brings 
\be
\mu_i \simeq \bar{\beta}^2 \left ( \frac{2 \dot M_w}{\dot M_{ai}}\right ) \left (  \frac{u_o}{\Omega_Kh} \right )_i \frac{r_i^2}{\Delta r_i^2} = \bar{\beta}^2 \frac{f}{1-f} m_{s,i} \frac{r_i^2}{\Delta r_i^2}
\label{eq:mui}
\ee 
which is general, valid for any fan-shaped wind configuration. In this expression, we introduced $f= \frac{2 \dot M_w}{\dot M_{ae}}$ as the ratio of the total (two-sided) mass loss in winds to the incoming mass rate in the disc at $r_e$.
 
The angular momentum conservation in the jet writes $L(\Phi)= \Omega_*r^2_A = \Omega r^2 - rB_{\phi}/\eta$ where $r_A$ is the Alfv\'en radius and $L(\Phi)$ the total specific angular momentum carried along a magnetic surface. Normalizing it to $\Omega_{K,i} r_i^2$ and evaluating it at the disc surface gives
\be
J(\Phi) =\frac{L(\Phi)}{\Omega_{K,i}r_i^2}\simeq 1 + \frac{q}{\kappa_{BP}} \simeq 1 + \beta \bar \beta q \frac{r_i}{\Delta r_i}
\ee
which is valid for any fan-shaped wind configuration. Note that $J= \lambda_{BP}$, the \citet{blan82} magnetic lever arm parameter.  

The projection of the momentum conservation equation along a magnetic surface provides the Bernoulli invariant
\be
E(\Phi)= \frac{u^2}{2} + \Phi_G -\Omega_*\frac{rB_{\phi}}{\eta} = {\cal E}(\Phi) + \Omega_*^2r^2_A  \ ,
\ee 
which is the total specific energy carried by each magnetic surface. However, only a fraction ${\cal E}(\Phi)$ is actually available for accelerating the plasma, the part $\Omega_*^2r^2_A$ (the energy of the magnetic rotator) remaining unused. If, at infinity, the magnetic field structure keeps almost no energy, then the flow reaches its maximum allowable speed $v_{max}(\Phi)=  \Omega_{Ki}r_i ( 2 J(\Phi) - 3)^{1/2}$. 

The total power carried away by the cold wind through the disc surface  is
\be
2P_{\mathrm{wind}}= 2 \int_{} E(\Phi)\rho {\bf u}_p \cdot {\bf dS} =  2 P_{\mathrm{MHD}} + 2 P_{\mathrm{kin}}
\ee
The kinetic power carried by the outflowing plasma at the disc surface, namely
 \be
P_{\mathrm{kin}} = \displaystyle\int \left(\frac{u^2}{2}+\Phi_G\right) \rho{\bf u}_p  \cdot  {\bf dS} 
\simeq -f\frac{GM\dot{M}_{ae}}{4r_i}
\ee
is initially negative: the wind power is mostly stored as magnetic power that will be eventually transferred to the plasma. The dominant contribution in $P_{\mathrm{wind}}$ is the MHD Poynting flux ${\bf S}_{MHD} $ leaving the disc
\begin{eqnarray}
 P_{\mathrm{MHD}}&=& \displaystyle\int - \Omega_* r \frac{B_\phi^+}{\mu_o} {\bf B_p} \cdot {\bf dS} \simeq 
  - \left . \frac{B_\phi^+ B_z}{\mu_o} \right |_i 2 \pi r_i^2 \Omega_K \Delta r_i 
 \nonumber \\
& \simeq & \frac{GM \dot{M}_{ai}}{4 r_i} \left . \frac{ 2 q \mu}{m_s} \right |_i \frac{\Delta r_i}{r_i}
\end{eqnarray}
Inserting Eq.~(\ref{eq:mui}) into this expression provides a wind power
\begin{eqnarray}
2P_{\mathrm{wind}} &\simeq &\frac{GM \dot{M}_{ae}}{2 r_i} \left ((1-f) \left . \frac{ 2 q \mu}{m_s} \right |_i \frac{\Delta r_i}{r_i}- f \right ) \nonumber \\
&\simeq &  \frac{GM \dot{M}_{ae}}{2 r_i}  \left ( 2q \bar \beta^2 f \frac{r_i}{\Delta r_i} - f \right ) 
\label{eq:Pwind}
\end{eqnarray}
Defining the average maximum jet velocity $v_w$ as $P_{\mathrm{wind}} = \frac{1}{2} \dot M_w v_w^2$, one obtains 
\be
v_w = \Omega_{Ki}r_i \left( 2q \bar \beta^2\frac{r_i}{\Delta r_i} - 1\right)^{1/2}
\ee
This expression is nothing more than $v_w = \Omega_{Ki}r_i (2 \bar J - 3)^{1/2}$, with an average magnetic lever arm $\bar J= 1 + q \bar \beta^2 r_i/\Delta r_i$. This result is again general and valid for any cold fan-shaped wind arising from a near-Keplerian accretion disc. 

\subsection{The global energy budget of the disc}

The local energy conservation law for a steady-state MHD flow is given by the following expression
\be
{\bf \nabla}\cdot\left(\rho{\bf u}_p\left[\frac{u^2}{2}+\Phi_G \right] + {\bf S}_{MHD} + {\bf S}_{rad}- {\bf u}\cdot{\mathbf T}\right) = 0
\label{eq:cons}
\ee
where $\Phi_G, {\bf S}_{MHD}, {\bf S}_{rad}, \mathbf T$ are respectively the gravitational potential of the central star, the MHD Poynting flux, the radiation flux and the turbulent "viscous" stress tensor (${\bf u}$ stands for the velocity of the fluid and $\rho$ for the plasma density). The last term in Eq.(\ref{eq:cons}) mimics  the energy transport by turbulence and vanishes at the disc surface. The plasma enthalpy may be neglected here. Using the Green-Ostrogradky theorem on the volume defined by the disc (actually an annulus of extent $\Delta r_i$ and height $h$) allows to derive the global energy budget, namely
\be
P_{\mathrm acc} + P_{\mathrm vis} = 2P_{\mathrm wind} + 2P_{\mathrm rad}
\ee
In this equation, the total power carried away by the cold winds $2P_{\mathrm wind}$ (sum of kinetic and magnetic energy fluxes) and the disc luminosity $2P_{\mathrm rad}$ are fed by the power available in the disc between $r_i$ and $r_e$. This power has two possible origins: the released accretion power $P_{\mathrm acc}$ and the power $P_{\mathrm vis}$ brought in by viscosity. Any other energy flux, like e.g. heat deposition by irradiation, is assumed negligible here. 

\subsubsection{The accretion power}
By definition, the accretion power is the mechanical power released by the plasma accreting from the external radius of the launching zone $r_e$ to the inner radius $r_i$, namely
\begin{eqnarray}
P_{\mathrm acc} &= &- \left [ \int_{-h}^{+h}\left(\frac{u^2}{2}+\Phi_G \right)  2 \pi r \rho u_r  dz \right ]^{r_e}_{r_i} \nonumber \\
&= &- \left [  \frac{GM\dot{M}_{a}}{2r} \right ]^{r_e}_{r_i} 
\end{eqnarray}
where $[A]^{r_e}_{r_i}= A(r_e)-A(r_i)$. Because of the wind, the disc accretion rate $\dot M_a(r) =  - \int_{-h}^{h} 2 \pi r \rho u_r  dz$ is a function of the radius. Defining $\dot M_{ai} = \dot M_a(r_i)$ and $\dot{M}_{ae}=  \dot M_a(r_e)$, mass conservation writes $ \dot{M}_{ai} =\dot{M}_{ae} - 2 \dot{M}_{w}= \dot{M}_{ae} (1-f)$, leading to 
\be
P_{\mathrm acc} = \frac{GM\dot{M}_{ae}}{2r_i}\left( \frac{\Delta r_i}{r_i} - f \right)
\label{eq:Pacc}
\ee
It can be readily seen that there is no mechanical power available if  $ f > \Delta r_i/r_i$: there is not enough radial contrast to feed large amounts of mass with mechanical power.

\subsubsection{The "viscous" power}
In a turbulent accretion disc, the inward flux of mass coexists with outwardly directed fluxes of energy and angular momentum. At each radius, energy and angular momentum are therefore being deposited from the inner radii, which translates here into a net power
\begin{eqnarray}
P_{\mathrm vis} &=&  \left [ \int ({\bf u}\cdot{\mathbf T})\cdot{\bf dS} \right ]^{r_e}_{r_i} = \left [ \int_{-h}^{+h} 2 \pi r \Omega r T_{r\phi} dz \right ]^{r_e}_{r_i} \nonumber \\
&= &- \left [  \frac{3}{2} {\cal R}_e^{-1} \, \frac{GM\dot{M}_{a}}{r} \right ]^{r_e}_{r_i}  \nonumber \\
& \simeq & \frac{3}{2} {\cal R}_{e,e}^{-1} \, \frac{GM\dot{M}_{a,e}}{r_i} \left ( {\cal D} + \frac{\Delta r_i}{r_i} \right )
\end{eqnarray}
where 
\be
{\cal D} = (1-f)\frac{{\cal R}_{e,e}}{{\cal R}_{e,i}} -1 =  \frac{\Sigma_i}{\Sigma_e} \frac{\nu_{v,i}}{\nu_{v,e}} - 1
\label{eq:defD}
\ee
Since the only available sources of energy are $P_{\mathrm acc}$ and $P_{\mathrm vis}$, the difference factor ${\cal D}$ needs to be positive or null. When no wind is present, the power released by accretion accounts  for only one third or less of the power available, namely
\be
\frac{ P_{\mathrm vis}}{P_{\mathrm acc} } = \frac{3}{{\cal R}_{e,e}}\left( {\cal D}\frac{r_i}{\Delta r_i} + 1\right) =  2 \left ( 1 +{\cal D} \frac{r_i}{\Delta r_i} \right )
\ee
where the value of ${\cal R}_{e,e}$ is set to $3/2$ to match the physical properties of the external SAD. Thus, energy deposition by viscosity (the value of ${\cal D}$) is of critical importance only for a small radial extent\footnote{In standard accretion disc theory, the power lost at the external radius by viscosity is negligible with respect to the accretion power $P_{\mathrm acc}$ because $r_e \gg r_i$. On the other hand, the "zero torque condition" (i.e. $T_{r\phi}=0$) assumed at the inner boundary $r_i$ forbids any incoming flux of energy. Both assumptions ($r_e \gg r_i$ and $T_{r\phi}=0$) are violated in the context of fan-shaped winds.} $\Delta r_i\ll r_i$. We will come back to this crucial issue later.

\subsubsection{The disc luminosity}
The luminosity $2P_{\mathrm rad}$ of a thin disc is the power that has been dissipated into heat by turbulence inside the disc and radiated away at its surfaces. In a steady-state framework it writes
\be
2P_{\mathrm{rad}} = \int_V dV \left [  \rho \nu_{\mathrm v} \left ( r \frac{\partial \Omega}{\partial r} \right)^2 + \eta_{\mathrm m} J_\phi^2 + \eta'_{\mathrm m} J_p^2  \right ] 
\ee 
where the integration is made over the volume $V$ of the disc. We can rewrite $2P_{\mathrm{rad}} =   \int_{r_i}^{r_e} Q(r) dr$ where $Q = Q_v + Q_m$, namely  the viscous term is
\be
Q_v(r) =\frac{GM \dot M_a}{r^2}   \frac{9}{4{\cal R}_e} 
\ee
and the magnetic (effective Ohmic heating) is
\be
Q_m(r) =    \frac{GM \dot M_a}{r^2} \mu {\cal F}
\ee
with ${\cal F}= {\cal R}_m \varepsilon^2 + \displaystyle\frac{q^2}{{\cal R}_m \chi_m}$ and  ${\cal R}_m = {\cal R}_e {\cal P}_m $ is the magnetic Reynolds number (vertically averaged). The factor $ \mu {\cal F}$ is estimated to be at most of the order of $\varepsilon=h/r$ in all situations. As a consequence, the magnetic dissipation term is always negligible with respect to the viscous dissipation.  Since we are dealing with a thin annulus where most of the dissipation is expected to occur at the innermost radius, we derive the following estimate for the disc luminosity
\be
2P_{\mathrm{rad}} \simeq Q_v(r_i) \Delta r_i =  \frac{9}{2}  {\cal R}_{e,e}^{-1} \frac{GM \dot M_{a,e}}{2r_i}  \frac{\Delta r_i}{r_i} (1 + {\cal D} )
\ee
It is interesting to compare this power to the incoming one by viscosity, namely
\be
\frac{ 2P_{\mathrm rad}}{P_{\mathrm vis} } \simeq \frac{3}{2} \displaystyle\frac{\Delta r_i}{r_i} \frac{ 1 + {\cal D}}{  \frac{\Delta r_i}{r_i} + {\cal D} } 
\ee
In the absence of any ejection ($f=0$), the energy balance of the annulus would be simply $P_{\mathrm acc} + P_{\mathrm vis} = 2P_{\mathrm rad}$, which provides ${\cal D} =0$. In order for winds to be present ($f \neq 0$), the difference factor  ${\cal D}$ must be positive.

\subsubsection{The power available for winds}
\label{sec:Pwind} 
The available power feeding cold fan-shaped winds is
\begin{eqnarray}
2P_{\mathrm wind} &=& P_{\mathrm acc} + P_{\mathrm vis} - 2P_{\mathrm rad} \nonumber \\
&\simeq&  \frac{GM \dot M_{a,e}}{2r_i} \left ( 
 \frac{\Delta r_i}{r_i} - f +  \frac{3} {{\cal R}_{e,e}} \left ({\cal D} +  \frac{\Delta r_i}{r_i} \right ) \right . \\
& & - \left .  \frac{9}{2} \frac{\Delta r_i}{r_i}  \frac{1+{\cal D}}{{\cal R}_{e,e}} \right ) \nonumber
\end{eqnarray}
where Ohmic dissipation has been neglected. Equating this estimate with expression (\ref{eq:Pwind}) and using the boundary condition  ${\cal R}_{e,e}= 3/2$ provides the following important constraint
\be
q \bar \beta^2 f  \simeq  {\cal D} \frac{\Delta r_i}{r_i}
\ee
that must be fulfilled for any kind of fan-shaped wind. This expression relates the mass loading parameter $\bar \beta$ of fan-shaped winds to disc physics ($q,f$) and in particular to the difference factor ${\cal D}$. The average magnetic lever arm of fan-shaped winds writes $ \bar J = 1 + {\cal D}/f$ while their power is
 \be
2P_{\mathrm wind} \simeq  \frac{GM\dot{M}_{a,e}}{2 r_i} \left ( 2 {\cal D} - f \right )
\label{eq:Pwin_D}
\ee
and is directly controlled by ${\cal D}$. The value of $f$ must be obtained by a complete calculation of the two-dimensional axisymmetric accretion-ejection problem. In the absence of such calculation, we use $f$ as a free parameter, simply fulfilling $f < 2 {\cal D}$. {\em It is clear that, unless ${\cal D}$ is of order unity or larger, fan-shaped winds are just epiphenomena with respect to accretion, both in terms of power and mass loss rates}. As a consequence, this implies that the power feeding such fan-shaped winds must be entirely supplied by viscous transport from the regions below $r_i$. It is important to remember this point.

\section{The various kinds of fan-shaped winds}

In the previous section we investigated some generic properties of the disc and its wind, regardless of the configuration. We just assumed the existence of steady-state cold fan-shaped winds launched from a disc annulus of extent $\Delta r_i \sim h$. This annulus is settled between an inner radius $r_i$ where the poloidal magnetic field is near equipartition ($\mu_i \sim 1$) and an outer radius $r_e$, where a standard accretion disc is established with $\mu_e \ll 1$ and ${\cal R}_{e,e}=3/2$. In the following, we analyze the three possible configurations envisioned in the introduction and discuss also the particular case of the X-wind.  

\subsection{T-winds}

A Terminal-wind is located at the outer part of a JED, a disc which is launching cold jets. By definition, one expects $p= B_r^+/B_z$ of order unity and thus ${\cal R}_{m,i} \simeq p r/h \sim \varepsilon^{-1}$ (see Appendix \ref{AppA}). As a consequence, the Reynolds number at the JED interface is ${\cal R}_{e,i}= {\cal R}_{m,i}{\cal P}^{-1}_m \simeq {\cal P}^{-1}_m r/h$, which translates into 
\be
{\cal D} \simeq \frac{3}{2}\frac{h}{r}(1-f){\cal P}_m - 1
\ee
A turbulent disc with ${\cal P}_m \sim 1$ would display ${\cal D}\simeq -1$ and the situation considered here would be energetically impossible. The reason is quite obvious: the annulus would loose much more energy at $r_e$ by viscous friction than what it would gain at $r_i$, even considering the case $f=0$. Thus, either the inside-out transition from a JED to a SAD is done over a larger extension, namely $\Delta r_i \sim r_i$ or more (but then the assumption of winds taking the shape of a fan would break down), or  ${\cal P}_m$ is larger than $r/h$ so that ${\cal D}>0$ becomes non negligible.

This has however strong implications on the inner JED physics. Indeed, a JED requires $m_{s,i}^{mag} \simeq \alpha_{m,i}$ of order unity \citep{ferr97,zann07}, which implies that the ratio of the magnetic to the viscous torque becomes 
\be
\Lambda_i = \left. \frac{m_s^{mag}}{m_s^{vis}}\right |_i \simeq \left . \frac{1}{\zeta  {\cal P}_m \varepsilon} \right |_i \sim \frac{1}{\zeta_i} \sim 1
\ee
since $\zeta= 3/2 +3\xi$ in a JED where $\dot M_a \propto r^\xi$. Thus, ${\cal P}_m\sim r/h$ or larger implies a viscous torque that remains comparable to the magnetic torque in the whole JED zone. This is why some power can be successfully conveyed outwardly by viscosity and ultimately feeding the T-winds. Defining $r_{in}$ as the innermost radius of the JED, this power can be estimated as
\be
\frac{2P_{\mathrm wind}}{P_{\mathrm acc, JED}} \simeq\left ( \frac{r_{in}}{r_i} \right )\frac{(2D-f)}{1-f}
\ee 
where $P_{\mathrm acc, JED}$ is the available power for the JED and $r_i$ corresponds to its outer radius ($r_J$, Fig.~1). If  ${\cal P}_m \sim \varepsilon^{-1}$ is such that $ {\cal D}$ is of the order of $\frac{\Delta r_i}{r_i}$ only, then T-winds would be just epiphenomena  located outside a more powerful disc wind and allowing the transition from an inner JED to an outer SAD. This sounds quite reasonable and might deserve some attention. But if ${\cal P}_m$ is larger so that  $ {\cal D}$ if of order unity then the inner JED undergoes a significant loss of power by friction at its outer radius and one cannot neglect this loss in the global energy budget of JEDs. 

However, regardless of these considerations, ${\cal P}_m \sim \varepsilon^{-1}$ or larger would require the following turbulence parameters in the disc
\begin{eqnarray}  
&& \alpha_{v,i} \sim \varepsilon^{-1} \nonumber \\
&& \alpha'_m \sim 1 \label{eq:turbTwind}\\
&& \alpha_m \sim 1 \nonumber
\end{eqnarray}
Obtaining such a large viscosity is quite problematic and is probably an indication that one or more assumptions are actually violated.

\subsection{M-winds below $R_{co}$}

When the stellar magnetic field is such that $R_t < r_i < R_{co}$, namely a disc truncation below the co-rotation radius, accretion onto the star is able to proceed along magnetospheric curtains (accretion columns or funnels). This is possible because the stellar magnetic field brakes down the disc material below $r_i$, in a way that is pretty much the same as in a JED. While the stellar magnetosphere is strong enough to truncate the disc right below $r_i$ with $\mu_i $ of order unity, it efficiently brakes down the disc material giving birth to a sonic accretion flow at $r_i$ so that an accretion funnel flow can take place \citep{roma02, roma09, bess08, zann09}. We identify therefore $r_i$ as the base of the funnel flow (termed $r_{bf}$ in \citealt{bess08}). This situation is therefore very similar to the T-wind case above, where the two torques (viscous and magnetic) are braking down the disc plasma.

If cold winds are to be launched between $r_i$ and $r_e$, then the poloidal magnetic field must be bent enough so that ${\cal R}_{m,i} \sim r/h$, leading to a Reynolds number ${\cal R}_{e,i} \sim {\cal P}^{-1}_m r/h$ as in the case of T-winds. The conditions for these M-winds are therefore exactly the same as for T-winds, namely $ {\cal P}_m > r/h$. Since $m_{s,i} \sim 1$ here also, the turbulence parameters must also verify (\ref{eq:turbTwind}).

\subsection{M-winds beyond $R_{co}$}

When the stellar magnetic field is such that  $R_{co} < R_t < r_i$, then no accretion columns can be formed. Indeed, the magnetosphere below $r_i$ is actually accelerating azimuthaly the disc material, which is pushed outwardly. In other words, the magnetic stresses due to the closed magnetospheric field lines deposit stellar angular momentum into the disc and accretion cannot proceed anymore below $r_i$. A steady-state configuration can be maintained only under two additional conditions:
\begin{enumerate}
\item Mass conservation requires the complete deviation of disc material into the M-winds, namely $f =1$.
\item Viscous stresses must be able to transport  the stellar angular momentum radially beyond $r_i$. This translates into a viscous torque that must be positive between $r_i$ and $r_e$.  If it were not for the torque due to the M-wind, accretion would stop. Assuming that the disc recovers a standard structure at $r_e$ demands a transition zone where the negative torque due to the M-wind is larger than the positive torque due to the viscosity.
 \end{enumerate}

Since there is no mass accreting below $r_i$, the power released by accretion is negative
\be
P_{\mathrm acc} = \frac{GM\dot{M}_{ae}}{2r_i}\left( \frac{\Delta r_i}{r_i} - 1 \right) \simeq -  \frac{GM\dot{M}_{ae}}{2r_i}
\ee
and one gets simply 
 \be
2P_{\mathrm wind} \simeq  \frac{GM\dot{M}_{a,e}}{2 r_i} \left ( 2 {\cal D} - 1 \right )
\ee
or, equivalently, winds with an average magnetic lever arm $\bar J= 1 + {\cal D}$ with $ {\cal D} > 1/2$. The power feeding the jets is here provided by the viscosity but its source is the rotational energy of the star.

The difference factor can also be written
\be
{\cal D} = \frac{\Sigma_i}{\Sigma_e} \frac{\alpha_{v,i}}{\alpha_{v,e}} - 1
\ee
which is always valid, including in the case $f=1$ considered here. The presence of mass $\Sigma_i$ stuck at $r_i$ (with a vanishing radial velocity) is required for the viscous transfer of energy from the rotating magnetosphere below $r_i$ to the annulus of extent $\Delta r_i$. M-winds of this kind could therefore only exist if
\be
\frac{\alpha_{v,i}}{\alpha_{v,e}} > \frac{3}{2}  \frac{\Sigma_e} {\Sigma_i}
\ee
Clearly, $\Sigma_i$ cannot be much larger than $\Sigma_e$ otherwise the thin disc (near Keplerian) approximation would break down as this would generate a significant radial pressure gradient. It is therefore safe to argue that the viscosity coefficient $\alpha_{v,i}$ at $r_i$ must be at least comparable to the one at $r_e$. 

Mass flux conservation requires $ 2 \rho^+ u_z^+ \Delta r_i \sim \Sigma_e u_e$ in a steady state framework, where all accreted mass is being deflected into the winds. Given our geometry with $\Delta r_i \sim h$, this implies $u_e \sim u_z^+ \sim C_s$ in order to start a wind. {\em In other words, the disc plasma is entirely deflected in the vertical direction into a wind only if the incoming velocity is already close to the sonic speed.} If this were not the case, mass would simply accumulate at the magnetopause $r_i$ and the steady-state approximation would break down. Because $u_e= m^{visc}_{s,e} C_s = 3 \alpha_{v,e} \varepsilon C_s/2$, the condition $u_e \sim C_s$ cannot be fulfilled unless $ \alpha_{v,e}\sim \varepsilon^{-1}$.

In order to halt accretion at $r_i$ ($m_{s,i} \simeq 0$), the viscous torque must balance the magnetic torque due to the M-winds, namely be such that $\zeta_i \simeq - 2 q_i \mu_i/\alpha_{v,i} \varepsilon <0$. Using Eq.~(\ref{eq:zeta}) in the transition zone with $f=1$ gives
\be
\zeta_i \simeq  \frac{3}{2} - 3 \frac{r_i}{\Delta r_i} \left( \frac{ {\cal R}_{e,e}}{ {\cal R}_{e,i}} - 2\right )
\simeq  - 3  \frac{r_i}{\Delta r_i} \left( \frac{3 \alpha_{v,i} \varepsilon}{2 m_{s,i}} - 2\right) 
\ee
The fact that the Reynolds number decreases drastically when approaching $r_i$ is consistent with a viscous torque becoming positive. The sonic Mach number consistent with the disc angular momentum conservation writes
\be
m_{s,i} \simeq  \frac{3}{2} \frac{ \alpha_{v,i} \varepsilon} {2 + \frac{2}{3} \frac{\Delta r_i}{h} \frac{q_i \mu_i}{ \alpha_{v,i}}}
\ee  
in a region where $\Delta r_i \sim h$ and $\mu_i \sim 1$. If $ \alpha_{v,i} > q_i$  then $m_{s,i} \sim \alpha_{v,i} \varepsilon > \alpha_{v,e} \varepsilon \sim1$. This is inconsistent with an arrested accretion. The only alternative is therefore to have $ \alpha_{v,i} \ll q_i$ which provides $m_{s,i} \sim \alpha^2_{v,i} q_i^{-1} \varepsilon$ that can be indeed vanishingly small. However, this requires $q_i \gg \varepsilon^{-1}$, namely magnetic field lines dramatically twisted. Note that it is because the viscous torque is huge at $r_e$ that the toroidal magnetic field must be so large.

The turbulence parameters must verify in this case
\begin{eqnarray}  
&& \alpha_{v,i} \ga \alpha_{v,e} \sim \varepsilon^{-1} \nonumber \\
&& \alpha'_{m,i} \ll \varepsilon \label{eq:turbMwind}\\
&& \alpha_{m,i} =  \alpha_{v,i} {\cal P}_m^{-1}\nonumber
\end{eqnarray}
which is again troublesome, regardless of the value of ${\cal P}_m$.

\subsection{M-winds from $R_{co}$: X-winds}

The limiting case when $R_t < r_i \simeq R_{co} $ is interesting because it may possess the good sides of the above two M-wind types without their drawbacks. Such a situation is commonly known in the literature as X-winds. First, some rotational energy might flow from the star to the disc beyond $r_i$ thanks to the viscosity (see discussion p786, left column, last paragraph in \citealt{shu94a}). Second, the ejected fraction $f$ could remain smaller than unity because the disc below $r_i$ would have the ability to form accretion columns. Let us then investigate this situation. 

The expressions given in Sect.~(\ref{sec:Pwind}) are valid in the case of X-winds as well and can be cast into the following form: 
\be
 \frac{\bar J - 1}{\bar \beta^2}  = q  \frac{r_i}{\Delta r_i} = \frac{{\cal D}}{ \bar \beta^2 f} 
 \label{eq:bilan}
\ee
with the constraint  $f < 2 {\cal D}$. So far, $f$ (or ${\cal D}$) is unspecified and free as long as a full MHD calculation including the disc remains undone. However, actual explicit calculations of X-winds do exist and provide useful informations \citep{shu94b, naji94, shu95, shan98, shan02, cai08}. These computed 2D flow structures are fan-shaped cold winds expanding from a point source with an initially super slow-magnetosonic speed, matched to an asymptotic jet cylindrical solution. 

All published solutions display $\bar \beta$ of order unity (between 1 and 3) and $\bar J$ between 2 and 7 (see e.g. Table~4 in \citealt{naji94} and Table~3 in \citealt{cai08}). As the underlying disc is a mere boundary condition, some assumption has been made in order to derive the value of the ejected mass fraction $f$. Indeed, it has been assumed that X-winds transport away the exact amount of disc angular momentum accreted onto the star, namely $\dot M_{ai} \Omega_i r_i^2= 2 \dot M_w \bar J \Omega_i r_i^2$. This translates into $f \bar J = 1-f$ (Eq.~(4.7a) in  \citealt{shu94a} gives a more complete relation, but published models satisfy this last scaling) or ${\cal D}= 1- 2f$ with $f$ ranging from 0.1 to 0.4 (Table~5 in \citealt{naji94}). Thus, X-winds as they appear in the literature require 
\begin{eqnarray}
 {\cal D}= 1-2f  \sim 1  & \mbox{and} & q \sim \frac{\Delta r_i}{r_i}
\end{eqnarray}
Note that these values are consistent with those found in the aforementioned papers: see for instance Eq.(4.19) in  \citet{shu94a} providing the power carried by the wind and the value assumed for $q$ in \citet{naji94} (see also discussion \S 2.6 in \citealt{shu94b}).  

Now, Eq.(\ref{eq:zeta}) writes for any $f <1$
\be
\zeta_i =  \frac{3}{2} - 3 \frac{r_i}{\Delta r_i}\left(\frac{{\cal D}}{1-f} - f\right) \sim - 3 \frac{r_i}{\Delta r_i}
\ee
which shows that the viscous torque is indeed positive as in M-winds beyond $R_{co}$. From the definition of the difference factor ${\cal D}$ one gets using ${\cal R}_{e,e}=3/2$
\be
{\cal R}_{e,i} = \frac{3}{2} \frac{ 1-f}{1 + {\cal D}} = \frac{m_{s,i}}{\alpha_{v,i} \varepsilon}= \frac{3}{4}
\label{eq:Rei}
\ee
showing that the sonic Mach number at $r_i$ must be $m_{s,i} =  2q_i \mu_i + \zeta_i  \alpha_{v,i} \varepsilon= 3 \alpha_{v,i} \varepsilon/4$ for $ {\cal D}= 1-2f$. With the constraints that both $\mu_i$ and ${\cal D}$ being of order unity and $q \sim \varepsilon$, one gets
\begin{eqnarray}  
&& \alpha_{v,i} \sim \varepsilon \nonumber \\
&& \alpha'_m \sim  \varepsilon^{-1} \label{eq:turbXwind}\\
&& \alpha_m \sim \varepsilon^2 \nonumber
\end{eqnarray}
where the last condition has been derived using ${\cal R}_m \sim \varepsilon^{-1}$ (cold ejection). These transport coefficients are those required within the disc so that X-winds can indeed be launched with $ {\cal D}= 1-2f  \sim 1$. The almost perfect matching between the two modes of angular momentum transport (turbulent/radial, wind/vertical) gives rise to a slight unbalance, allowing henceforth accretion.

\section{Discussion}

We have shown that, under reasonable circumstances, all fan-shaped winds (T-winds and all kinds of M-winds, including X-winds) put severe constraints on the underlying MHD turbulence. It appears that our conventional view of turbulence can be hardly reconciled with any kind of fan-shaped winds from thin (Keplerian) discs.  Before discussing further the consequences of our findings, let us first address some caveats of our analysis.

\subsection{Caveats of our analysis}

\subsubsection{MHD turbulence and anomalous transport}

In order to design steady-state models of accretion discs, one needs to take into account the turbulent transport of both angular momentum and magnetic field. It has been assumed that turbulence leads to a local transport that has been included in our equations as an anomalous viscous torque and an anomalous resistivity term. 
Self-sustained turbulence in accretion discs is thought to be triggered by the magneto-rotational instability \citep{balb91}. Thorough numerical and analytical analysis tend to show that this instability does indeed provide an outwardly directed angular momentum transport that, indeed, can be crudely described by a {\rm local} turbulent viscosity in a vertically averaged disc $\nu_{\mathrm v}$ (see \citealt{balb03}). Many important results have been obtained in the context of shearing boxes (e.g. \citealt{pess07,lesu07} and references therein). One of these results concerns the existence of an anomalous magnetic field transport in MRI-driven turbulence \citep{guan09,lesu09,from09}. While previous works were only focused on analyzing the angular momentum transport, these studies investigated the magnetic field dissipation. They concluded that it could also be roughly described by some  {\rm local} Ohmic resistivity term  with an effective magnetic Prandtl number ${\cal P}_m$ of order unity \citep{lesu09}. 

Two words of caution though. First, these studies were done only with very weak fields, namely $\mu \ll 1$ and  it remains therefore to prove that the behavior and scalings found would remain valid for a magnetic field near equipartition. Second, it is not clear how the boundary conditions in shearing boxes affect the development of the turbulence. Fully resolved and converged 3D simulations are therefore needed to verify if some non local behavior is present or not (see e.g. \citealt{beck09,beck11}). Anyway, given our present knowledge, we believe that using the functional form of the viscous stress tensor for the turbulent transport with a constant  ${\cal P}_m$ throughout the disc remains a reasonable, albeit limited, approach.

\subsubsection{Cold wind approximation}

In our analysis, we neglected the enthalpy in the disc energy budget and assumed field lines inclined by more than 30$^o$ with respect to the vertical. Both assumptions are not crucial but allow a simpler treatment of the equations. The amount of enthalpy carried in by the flow (both in the jet and at $r_i$) remains negligible as long as the disc is geometrically thin. The thin disc approximation goes together with the near-Keplerian assumption and provides indeed a limitation to our analysis. This work must  then be seen as a first attempt to understand whether fan-shaped winds can be steadily launched from near Keplerian discs.

In fact, including thermal effects in the launching process should not change our findings. Indeed, what is critical here is the strength of the poloidal magnetic field and its geometry. The magnetic field strength is assumed to decay outwardly so that $\mu$ goes from about unity at $r_i$ to a tiny value at $r_e$. Correspondingly with this decay, the poloidal field experiences a quasi-spherical dilution so that $B_r^+ \sim B_z$ between $r_i$ and $r_e$ is a natural outcome due to the geometry. Maintaining this geometry despite the natural tendency of the field to diffuse outwardly is an open issue that we do not address here. We assume a fan-like geometry and look at the dynamical consequences on the disc.

One could argue however that, contrary to the T-wind case, M-winds are located just outside a closed magnetosphere. This means that below $r_i$ the magnetic field lines are actually directed towards the star. Thus, $J_\phi$ must undergo a steep transition from $J_\phi \simeq B_r^+/\mu_o h \sim B_z/\mu_o h >0$ between $r_i$ and $r_e$ and a negative value below $r_i$. The poloidal field lines must then be almost vertical somewhere around $r_i$, defining thereby a zone from where no cold ejection could take place (a "dead zone",  \citealt{shu94a}). However, this will affect only few field lines around the radius where $J_\phi$ vanishes and $r_i$ must then be seen here as the radius where the fan-like geometry has already provided the correct bending.

\subsection{Are stationary fan-shaped winds possible?}

\subsubsection{T-winds and M-winds below co-rotation}

Both situations are very similar and energetically forbidden for ${\cal P}_m \sim 1$. On the other hand, even the case ${\cal P}_m \sim \varepsilon^{-1}$ is problematic as it would require a huge viscosity coefficient, $\alpha_{v,i} \sim \varepsilon^{-1}$. This is clearly against our current knowledge of MHD turbulence, where $\alpha_v <1$. Nevertheless, it may be of some interest to make numerical experiments, with $\alpha$ prescriptions and these parameters, to see whether fan-shaped winds would be launched. 
 
Our guess is however that our assumption of the very existence of fan-shaped winds is violated. Precisely, that no JED-SAD or Magnetosphere-SAD transition giving rise to winds can be done in a radial extent $\Delta r_i \sim h$. But since transitions like those are anyway necessary, we conjecture that they must be done on a larger scale  $\Delta r_i \sim r_i$ for a steady-state to settle.

\subsubsection{M-winds beyond co-rotation}

This configuration is even more problematic. It not only requires  $\alpha_v \sim \varepsilon^{-1}$ as the other two, but a steady ejection of all the incoming mass requires a huge magnetic shear $q \gg \varepsilon^{-1}$. It is doubtful that such a large winding of the magnetic field lines could be steadily maintained. Reconnection will probably take place so as to (violently) relax the magnetic field configuration to a lower energy state. Alternatively, for instance if $\alpha_v < 1$, then the disc material accreting below $r_e$ would be unable to be vertically deflected with $f=1$ and mass would start to accumulate. 

Our guess is that steady state M-winds will never be established beyond $R_{co}$.

\subsubsection{X-winds}

From the above discussion it appears that the only remaining configuration that could lead to stationary fan-shaped winds is when $r_i$ is near the co-rotation radius, namely X-winds. As already stated, we recovered with our formalism the main results of X-winds concerning the strength of the magnetic field ($\mu_i\sim 1$, \S 2.8 in  \citealt{shu94a}), the value of the magnetic twisting $q \sim \varepsilon$, the link between the MHD invariants ($\bar \beta, \bar J$) and the global energy budget (${\cal D}, f$). There are however two main differences.

First,  \citet{shu94a} do not use the usual single fluid MHD induction equation for the magnetic field. Instead, ambipolar diffusion is assumed to be the main source of magnetic diffusion, with a possible comparable role of the Ohmic resistivity (see their \S 3.1). However, the ambipolar diffusion term can be translated {\it in order of magnitude} into a diffusivity
\be
\nu_{AD} \sim \left ( \frac{\rho_n}{\rho} \right )^2 \frac{V_A^2}{\nu_{ni}}
\ee
where $\nu_{ni}$ is the neutral-ion collision frequency and $\rho\simeq  \rho_n + \rho_i = \rho_n (1+ X)$ with $\rho_n$ ($\rho_i$) the neutral (ion) density and $X$ the ionization fraction \citep{ferr93a}. Normalizing this diffusivity by $\Omega_K h^2$ gives another alpha coefficient, measuring the level of the ambipolar diffusion
\be
\alpha_{AD} \sim \frac{\mu}{(1+X)^2} \Omega_K \tau_{ni}
\ee
with $\tau_{ni}= \nu_{ni}^{-1}$. It is clear that a fluid approach requires $\Omega_K \tau_{ni} \ll 1$ within the disc and it is indeed the assumption made in X-wind theory.  However, this implies $\alpha_{AD} \sim \mu \Omega_K \tau_{ni} \ll \mu \la 1$, much smaller than the value $\alpha'_m$ required to maintain $q \sim \varepsilon$. We thus believe that, as long as some MHD turbulence is triggered and sustained within the inner disc, it will be the dominant diffusion term (note also that in the ambipolar diffusion regime, no MRI would be permitted, \citealt{bai11}).  

A second difference is in the use of the disc angular momentum. While we explicitly used it to compute the sonic Mach number $m_s$ (Sect~\ref{sec:mom}), we found only an order of estimate in  \citet{shu94a}, namely $m_s \simeq \Omega_K \tau_{ni} \ll 1$. However, the actual {\it value} of $m_s$ remains unspecified. Within our framework and with ${\cal D}$ of order unity, we obtain $m_{s,i} \sim \varepsilon^2$, which is indeed accretion done at a very slow pace. This is understandable as the angular momentum removal by the wind has to overcome the viscous deposit. But this is however problematic as the formation of accretion columns does require an almost sonic accretion at the base of the funnel flow \citep{bess08}. How will the disc material be loaded onto the magnetospheric field lines below $r_i$ remains an unsolved issue within the X-wind scenario.

Another drawback concerns the value of the turbulence parameters that are required for powering X-winds with ${\cal D}\sim 1$ in a steady state. The low accretion speed combined with the small magnetic shear require (see Eq.\ref{eq:turbXwind}) 
\be
\chi_m {\cal P}_m \sim \varepsilon^2
\ee
namely a huge anisotropy in magnetic field dissipation, regardless of the value of ${\cal P}_m $. This is inconsistent with our current knowledge of turbulence in shearing boxes.

\subsubsection{Designing different X-wind solutions}

The double goal sought by the X-wind model, as theorized in  \citet{shu94a}, is to explain YSO jets while simultaneously braking down the protostar. More precisely, X-winds should be the dominant jet (in terms of appearance, mass flux and power) while carrying away enough angular momentum so that the star is not being spun up by the accreting material (zero net torque condition). These requirements lead to the choices of ${\cal D}= 1-2f \sim 1$, $q\sim \varepsilon$, $\bar J$ of order of a few and $\bar \beta$ of order unity. However, these values raise the issues discussed above. 

The difficulties faced by these M-winds can be understood in a broader framework. As expressed by Eq.~(\ref{eq:defD}) ${\cal D}$ measures the net energy deposition by viscosity in the zone of extent $\Delta r_i \sim h$. This deposit will eventually power the jets (Eq.\ref{eq:Pwin_D}) and that is why M-winds of any kind require ${\cal D}$ of order unity to be of any significance. The problem is the tiny width of the launching zone, which imposes that $ {\cal D}\sim 1$ requires a Reynolds number of order unity as well. Otherwise there is much more power lost at $r_e$ than the one gained at $r_i$ and ${\cal D}\simeq -1$, Eq.(\ref{eq:Rei}). This translates unavoidably into $m_{s,i} \sim \alpha_{v,i} \varepsilon$ which is of order unity only for $\alpha_{v,i} \sim \varepsilon^{-1}$.

Let us design new M-wind solutions by demanding $ {\cal D}\sim 1$  and $m_{s,i} \sim 1$ only. $ {\cal D}$ of order unity requires $\zeta_i \sim - \frac{r_i}{\Delta r_i}$ and the viscous torque is accelerating the disc material. Then, to maintain accretion at a sonic pace, the magnetic torque must be correspondingly huge with $q_i \sim  \frac{r_i}{\Delta r_i}$. As a consequence, the interplay with the jet (Eq.\ref{eq:bilan}) imposes $\bar \beta \sim  \frac{\Delta r_i}{r_i}$ if one desires winds with asymptotic speeds comparable to those observed. In practice, this means that new super-fast MHD wind solutions must be looked for.
 
Nevertheless, even if such wind solutions could be found, they will demand the following $\alpha$ coefficients in the disc
\begin{eqnarray}  
&& \alpha_{v,i} \sim \varepsilon^{-1} \nonumber \\
&& \alpha'_{m,i} \sim  \varepsilon \label{eq:Xwind_best}\\
&& \alpha_{m,i} \sim 1 \nonumber
\end{eqnarray}
with both ${\cal P}_m \sim \varepsilon^{-1}$ and $\chi_m \sim  \varepsilon^{-1}$. This is again troublesome, not only because of the strange anisotropy that would be required but because of the overwhelming magnetic compression on the disc implied by $q\sim \varepsilon^{-1}$ with a near equipartition field: no vertical balance could be found (see Appendix \ref{AppB}).

Finding M-wind solutions that would be consistent with our conventional knowledge on MHD turbulence, namely ${\cal P}_m \sim \chi_m \sim 1$ is impossible. Indeed, ${\cal D}>0$ (even small) requires ${\cal R}_{e,i}$ of order unity while the geometry itself requires ${\cal R}_{m,i} \sim \varepsilon^{-1}$. Thus, the minimum requirement is ${\cal P}_m \sim \varepsilon^{-1}$ \citep{shu07}, namely $ \alpha_{m,i} \sim 1$ and $\alpha_{v,i} \sim \varepsilon^{-1}$. Now, if one requires that $q$ is at most of order unity  with ${\cal D}\sim 1$, then Eq.(\ref{eq:bilan}) requires $f \bar \beta^2 \sim \varepsilon$. Thus, new dynamical flow calculations should be looked for since, for $\bar J \simeq 1 + {\cal D}/f$ comparable to values used in current X-wind models, the mass loading parameter should be $\varepsilon^{1/2}$ times smaller. How this impacts on the wind dynamics needs to be investigated as $\bar \beta \ll 1$ describes a priori a matter dominated flow ($\bar \kappa_{BP}\sim 1$) and it is not clear that any super-Alfv\'enic wind could be obtained.

\subsection{Comparison with numerical simulations}

\subsubsection{T-winds}
There is no proper {\em ab initio} MHD simulation of turbulent accretion discs offering a radial transition from an inner JED to an outer SAD. Simulations with well defined JEDs are those done using $\alpha$ prescriptions \citep{cass02, zann07,tzef09,murp10}. However, in most simulations $\mu$ is a constant of the radius and the outer radius of the JED is continuously increasing in time. The only simulation with a steady-state super-fast jet launched from a zone of constant extent has been performed by \citet{murp10}, with an initial magnetization decreasing with the radius. 
Clearly, there is no fan-shaped wind making the transition from the inner zone with super-fast jets to the outer zone embedded in a  current-free magnetospheric corona (see e.g. Figs~1 and 2 in \citealt{murp10}). It might be because ${\cal P}_m$ is of order unity in these simulations: in that case, no sharp transition (i.e. $\Delta r_i\sim h$) with a spherical dilution can be energetically feasible. Instead, a zone of sub-fast but still super-Alfv\'enic jets is surrounding the inner zone from $r_i \simeq 5$ to $r_e = 13$, namely with an extent $\Delta r_i \sim r_i$. Alternatively, it may be just an effect of the initial magnetic field distribution, since the simulation lasted for 953 inner Keplerian orbits only, which is roughly the accretion time scale from $r=13$. Performing longer simulations is a task that should deserve some attention.

\subsubsection{M-winds}
MHD simulations of a star-disc interaction with an axisymmetric dipole field have been done by several authors, mostly using $\alpha$ prescriptions for the viscosity and magnetic diffusivities (see e.g. \citealt{roma02, roma09, bess08, zann09} and references therein). However, all of them were done with $\chi_m= 1$ and with both $\alpha_v$  and $\alpha_m$ smaller than unity. As a consequence, they are all outside the parameter range required to produce steady state M-winds as described in this paper, even if some of them were performed with ${\cal P}_m \sim \varepsilon^{-1}$ \citep{zann09,roma09}.   

\citet{zann09} showed that decreasing the value of $\alpha_m =\alpha'_m$ gives rise to an opening of the stellar magnetic field and the loss of any causal link between the star and the disc. This is because diffusion is not efficient enough and the star-disc differential rotation leads to a huge generation of toroidal field. The response of the magnetosphere to this increasing shear is to inflate and reconnect (\citealt{roma98} and discussion in  \citealt{matt05a}), breaking thereby the connection with the star. Depending on the parameters, some unsteady ejection events (plasmoids) can be both thermally and magnetically launched after a reconnection. But these plasmoids carry no self-confining electric current and have nothing in common with steady state M-winds. 
The existence of steady "conical winds" was however reported in the simulations of  \citet{roma09} when the disc is truncated below the co-rotation radius. Such winds seem to require  ${\cal P}_m \ga 1$ to exist but the mass loss $f$ decreases when ${\cal P}_m$ increases. Although the origin of these winds has not been fully addressed, they clearly do not resemble fan-shaped winds. Instead, they look better like usual disc winds that would be launched only from a small radial extent (the region where $\mu \la 1$) and confined by the presence of an outer open magnetic flux. When the disc is truncated beyond the co-rotation radius, the system enters the so-called "propeller regime" (\citealt{roma03a,roma05, usty06} and references therein), which is unsteady. 

{\em Ab initio} simulations that compute also the disc turbulence have been performed recently, both in 2D \citep{roma11} and 3D \citep{roma12}. These simulations are highly valuable as they are not biased by any $\alpha$ prescription. A first important result is that the disc truncation radius is indeed located at the position predicted using analytical arguments, namely where $\mu \sim m_s \sim 1$ (see \citealt{bess08}). A second result is that, in the 2D simulations where the effective magnetic resistivity seems to be very small (${\cal P}_m \gg 1$), still no M-winds were obtained. This holds true in 3D \citep{roma12}. This deserves further investigation, but the reason might be that a turbulent disc structure is unable to provide the required level of turbulent transport as measured by the conditions (\ref{eq:Xwind_best}).

\section{Conclusions}

We addressed in this paper the existence of cold, {\it steady-state}, fan-shaped winds arising from near Keplerian discs. Such winds are assumed to be launched from a disc annulus of radial extent $\Delta r $ of the order of the disc thickness $h$. They may play the role of a buffer between different magnetic regions in protostellar accretion discs. There are two main configurations (see Fig.~1). The first one, referred to as "Terminal-winds" (T-winds), acts as a transition layer between an inner Jet Emitting Disc and an outer standard accretion disc. The second kind of fan-shaped winds are the "Magnetospheric-winds" (M-winds), linking an inner closed  stellar magnetosphere to the outer standard accretion disc. Although described by the same set of MHD equations as usual disc winds, fan-shaped winds have interesting properties (such as the quasi-spherical dilution of the magnetic flux) that make them worth to be analyzed. \\ 

Taking into account  the physical conditions imposed by both inner and outer boundaries and considering the whole set of resistive, viscous MHD equations, we were able to describe for the first time the mass, energy and angular momentum balance of the near Keplerian disc. To drive winds with a significant power, the annulus must be fed at its inner edge with more energy than what is lost from winds, radiation and viscosity at its outer edge. We found that the only possible way to power such winds is to allow for an efficient {\it viscous energy flux} at the inner edge.  Such influx of energy by viscosity goes along with an influx of angular momentum. Thus, for accretion to take place despite this angular momentum excess, the torque due to the fan-shaped winds must be dominant wrt the local viscous stresses. Now, both the magnetic geometry and the balance of these angular momentum/energy fluxes put stringent constraints on the underlying MHD turbulence in the disc.\\ 

These conditions were analyzed using local $\alpha$ prescriptions for the transport coefficients: viscosity $\nu_v$, magnetic diffusivities $\nu_m$ and $\nu'_m$. We found the following results:

(1) For the case of T-winds and M-winds whose inner radius $r_i$ is located below the co-rotation radius $R_{co}$, we found that fan-shaped winds can exist only if the effective magnetic Prandtl number is larger than $r_i/h_i$ ($h_i$ being the disc thickness at $r_i$) and a viscosity $\nu_{v,i}$ larger than $\Omega_{K,i}r_ih_i$ ($\alpha_{v,i} > r_i/h_i$).  

(2) In the case of M-winds with $R_{co} > r_i$, we found a similar viscosity amplitude and a magnetic field configuration likely to be non-steady,  as it exhibits a huge magnetic twisting prone to reconnection. 

(3) We also considered in our study the particular case of M-winds whose internal radius coincides with $R_{co}$, namely X-winds as first proposed by \citet{shu94a}. We argue that, contrary to the assumption made in the earlier X-wind papers, ambipolar diffusion is probably not the dominant source of magnetic field diffusion at the star-disc magnetopause. Instead, since the large scale vertical magnetic field that is threading the disc must be close to equipartition, we expect strong field instabilities to be triggered there and to saturate into a self-sustained MHD turbulence (described here as local transport coefficients). The constraints imposed by all published super-Alfv\'enic X-wind flow models on the underlying disc structure lead to quite an extreme transport coefficient set-up. For instance, the degree of anisotropy of the magnetic field dissipation must scale as $\nu_m/\nu'_m \sim (h/r)^3$.\\

{\it None of the various transport coefficients set-ups obtained for T and M-winds are in agreement with our current knowledge of MHD turbulence in near Keplerian discs.} One key assumption made in this work is the small radial thickness $\Delta r \sim h$ of the wind launching zone. But if  $\Delta r \sim  r$, the geometry would not be consistent anymore with a spherical dilution of the magnetic flux, which is the cornerstone of  fan-shaped winds. Thus, unless MHD turbulence at a JED-SAD and Magnetosphere-SAD interface has quite peculiar properties, we believe that {\it steady-state} T-winds and M-winds of any kind (including X-winds) will not be realized in Nature.\\ 


While stationary T-winds and M-winds appear to be ruled out, would such type of outflows still be possible if the steady-state assumption is released ? Let us for instance consider that a thermal-viscous instability is triggered in the disc (see \citealt{laso01} for a thorough review on the Disc Instability Model). Its outcome is a thermal front, of width $\Delta r \sim h$, which propagates at almost sonic speeds, namely $u_{front}\sim \alpha_v C_s$, where $\alpha_v$ could be as high as 0.1 \citep{meye84, meno99}. Such a large accretion speed could advect large scale magnetic fields in the disc and give rise to a magnetic configuration as envisioned for T-winds. One could then wonder if fan-shaped winds just require specific conditions inside the disc, such as those necessary to trigger this instability. But under these circumstances, not only stationarity but also several other of our assumptions break down. Indeed, steep radial gradients of disc column density and temperature, both associated with this thermal front, can potentially change the rotation and specific angular momentum profiles so as to allow for a local Rayleigh instability \citep{lin85}. While our present analysis of fan-shaped winds took advantage of the existence of MHD invariants (and the strong links they provide between the winds and the disc), none of these would exist anymore in an outbursting disc prone to a thermal-viscous instability. Such links would disappear as the time scales involved (front propagation, wind readjustment) become too close. Note also that a magnetic configuration suited for M-winds could in principle be obtained in a time-dependent way. Indeed, it could be achieved if a large scale magnetic field advected towards the central object is suddenly halted by the magnetic pressure of the closed stellar magnetosphere and bounces back. These unsteady situations are however beyond the scope of our analysis, as full 2D time-dependent simulations would be needed to investigate such dynamical situations.\\

Actually, unsteady ejecta are indeed observed in many MHD simulations of star-disc interaction. Both in those computing the disc \citep{roma09, roma11} or where it is only a platform \citep{fend00,fend09}. As argued in \citet{ferr06b}, such ejecta are actually welcome. Indeed, they could produce time-dependent events that would be confined and channeled by an outer steady disc wind (driven from a JED), possibly explaining both jet variability and the presence of knots. Also, these ejecta could transport a sizable fraction of the stellar angular momentum and provide thereby a natural way to brake down the central protostar (see e.g. \citealt{ferr00, zann09}). According to this picture, there will be no unique MHD model capable of explaining all observational features of YSO jets. These would be instead a multi-component flow, made of a central core filled in by the stellar wind, an interface layer of unsteady plasmoids and an outer disc wind as the ultimate confining agent.

 \section*{Acknowledgements}
 We thank the anonymous referee for useful comments that draw our attention to the Disc Instability Model.


\appendix

\section{The poloidal magnetic field at the disc surface}
\label{AppA}
An axisymmetric magnetic bipolar topology is described by a poloidal field ${\bf B}_p  = \frac{1}{r}\nabla a \times {\bf e_{\phi}}$, where $a(r,z)$ is an even function of $z$, and an odd toroidal field $B_{\phi}$. A magnetic surface anchored at a radius $r_o$ is a surface of constant poloidal magnetic flux that can be labelled by $a(r,z)= a_o(r_o,0)$. The bending of the field lines at the disc surface ($z \sim h$) can be evaluated as 
\be
\frac{B_r^+}{B_z} \simeq h\frac{\partial^2 a_o}{\partial z^2}\times\left(\frac{\partial a_o}{\partial r}\right)^{-1}\sim \frac{h}{l^2} a_o\times\left(\frac{\partial a_o}{\partial r}\right)^{-1}
\ee
where $l$ is the vertical scale of variation of the magnetic flux. The radial distribution of magnetic flux through the disc is an unknown function. However, by definition, fan-shaped winds assume a steep decrease of $B_z$ with the radius, on a scale $\Delta r \sim h$. This is in strong contrast with extended disc winds as  $\displaystyle\frac{a_o}{\partial a_o/\partial r} \sim \Delta r \sim h \ll r$. As a consequence $B_r^+/B_z \sim h\Delta r/l^2 \sim h^2/l^2$. Extended disc wind configuration would lead to a vertical scale variation of the magnetic flux such that $B_r^+/B_z \sim hr/l^2$ \citep{ferr93a}.

Cold fan-shaped winds are launched by the very same magneto-centrifugal mechanism than large scale disc winds. They must thus fulfill the \citet{blan82} criterion. This geometrical criterion translates into $B_r^+/B_z$ of the order unity, which then implies $l \sim h$. Contrary to usual disc winds, $B_z$ varies on the disc scale height. However, this is not crucial for our estimates since $B_z^+= B_z(z=h)$ is smaller than but of the order of $B_z(z=0)$.

The cold, fan-like geometry of the magnetic field requires a toroidal electric current density at $r_i$ 
\be
J_\phi = \frac{1}{\mu_o} \left ( \frac{\partial B_r}{\partial z} - \frac{\partial B_z}{\partial r} \right ) \simeq c_1 \frac{B_z}{\mu_o h}
\ee
where $c_1>0$ is of order unity. According to Ohm's law (\ref{eq:diff}), the balance between field advection and diffusion produces
\be
J_\phi = - \left.\frac{r_i u_r}{\nu_m} \frac{B_z}{\mu_o r_i}\right|_{z=0} = {\cal R}_m \frac{B_z}{\mu_o r_i}
\ee
where  ${\cal R}_m$ is the effective magnetic Reynolds number. Thus, cold fan-shaped winds are possible only if  ${\cal R}_m \simeq r_i/h$.  This requires a quite significant accretion speed or, equivalently, a sonic Mach number $m_s = \alpha_m {\cal R}_m h/r \simeq \alpha_m$ at $r_i$.

\section{A near Keplerian quasi-static disc structure}
 \label{AppB}

The steady-state disc vertical equilibrium corresponds to a balance between three forces: gravity, magnetic pinching force and thermal pressure gradient. Integrating the z-component of the momentum conservation equation (\ref{eq:mom}) leads to 
\begin{eqnarray}
P_o - P^+ & \simeq & \frac{B_r^{+2}+B_{\phi}^{+2}}{2\mu_o} + c_2 \frac{B_z^2}{2\mu_o} \nonumber \\
& & +\,  \int_0^{h} \left(\rho z \Omega_K^2-u_z^2\frac{\partial \rho}{\partial z}\right)dz +\left[\rho\frac{u_z^2}{2}\right]_0^{h} 
\end{eqnarray}  
where $c_2>0$ is of order unity. Given the magnetic geometry, the magnetic Lorentz force is always negative, so that the only force allowing a vertical acceleration {\it within the disc} is the plasma pressure gradient. This is possible only if the magnetic pinching and compression are not too high, namely if 
\be
\frac{P_o - P^+}{P_o} \simeq 1 > \frac{\mu}{2} \left [ \left (\frac{B_r^+}{B_z}\right)^2 + \left (\frac{B_\phi^+}{B_z}\right)^2\right ]= \frac{\mu}{2} \left(p^2+q^2\right)
\ee  
where $p= B_r^+/B_z \geq 1$ and $q= -B_\phi^+/B_z$. This translates into an upper limit for the disc magnetization, namely $\mu < 2$ in order to satisfy the vertical equilibrium of the disc. The bending $p$ of the magnetic field is directly related to the magnetic Reynolds number through the resistive induction equation $p \simeq{\cal R}_m\varepsilon$ (see Appendix \ref{AppA}). A lower bound can then be obtained  for the magnetic Reynolds number, namely ${\cal R}_m \geq \varepsilon^{-1}$. Note that a steady-state situation is achieved only if $P^+ \ll P_o$, ie. $\rho^+ \ll \rho_o$. This simple argument shows why a M-wind in the propeller regime cannot be described within a steady-state framework.

In the zone of extent $\Delta r_i$, both the bending ($B_r^+/B_z$) and the shear ($B_\phi^+/B_z$) of the magnetic field lines increase with distance. But this can be easily compensated with the steep decrease in $B_z$ (hence $\mu$) so that the above condition can remain fulfilled. Note that, because of the vertical magnetic force, the real disc scale height is actually smaller than the hydrostatic scale height $h$, as defined in Eq.(\ref{eq:h}). However, this extra compression is at most of the same order than the gravitational one so that using $h$ and $C_s \simeq \Omega_K h$ is good enough for our purpose \citep{ferr95,shu08}. 

The steady-state angular velocity $\Omega$ of the plasma is given in the disc by the radial component of Eq.~(\ref{eq:mom}). Neglecting the advection term it writes on the disc midplane 
\begin{eqnarray}
\Omega_o^2 &\simeq &\Omega_K^2\left(1 + \frac{\partial P/\partial r}{\rho_o \Omega_K^2 r} - \frac{J_\phi B_z}{\rho_o \Omega_K^2 r}  \right) \nonumber \\
&\simeq & \Omega_K^2\left(1 + c_3 \frac{h^2}{r \Delta r_i}  - c_1 \mu \frac{h}{r} \right)
\label{eq:omega}
\end{eqnarray}
where $c_3$ is at most of order unity (both signs are allowed). The strong radial decrease of the vertical field gives a magnetic support against gravity.  However, for the conditions envisioned here, namely $\mu$ at most of order unity, a Keplerian rotation law remains a good approximation in a thin disc.

\section{The toroidal magnetic field at the disc surface}
\label{AppC}

The disc induction equation (\ref{eq:ind1}) for the toroidal component can be written as \citep{ferr95, ferr97}
\be
\eta'_mJ_r(z) = \left . \eta'_mJ_r\right |_{z=0} + r\int_0^z {\bf B}_p\cdot\nabla\Omega dz - B_{\phi}u_z
\label{eq:ind}
\ee
in the resistive MHD regime, where the last term (advection) can be neglected as the vertical velocity is very small compared to the toroidal velocity in a thin disc. At the disc surface the product $\eta'_mJ_r$ must vanish for two reasons. First, turbulence must decrease so that the flow becomes frozen in the magnetic field. Second, $J_r$ itself must also decrease to allow for a change of sign of the magnetic (wind) torque $F_\phi= J_z B_r - J_r B_z$. Indeed, from negative at the disc midplane, the torque must become positive at the disc surface. This is the cornerstone of magnetic launching from thin discs \citep{ferr93b,ferr95}. In the absence of dynamo, the only possibility is to balance the term $ \eta'_mJ_r$ at $z=0$ with the counter electromotive force (cemf) due to the disc differential rotation $r {\bf B}_p \cdot\nabla\Omega $ integrated over the disc thickness (see for instance Fig.~2 in \citealt{ferr93b}). 
Given the fan-shaped geometry, the main contribution to this cemf is the radial one, which writes $r \int_0^h B_r \frac{\partial \Omega}{\partial r} dz \simeq - \frac{3}{2} \Omega_K  \int_0^h B_r dz \simeq - c_4 \Omega_K  B_z h$, where $c_4 >0$ is of order unity for a cold wind.  
Thus, a vanishing $J_r$ at the disc surface is realized if, at the disc midplane, it amounts 
\be
J_{r,o} \simeq c_4 \frac{B_z}{\mu_o h} \left . \frac{\Omega_K h^2}{\nu'_m} \right |_{z=0} \simeq \frac{c_4}{\alpha'_m}  \frac{B_z}{\mu_o h} 
\ee 
Defining precisely the magnetic shear parameter $q$ as  
\be
J_{r,o}= q \frac{B_z}{\mu_o h} = - \left.\frac{1}{\mu_o}\frac{\partial B_\phi}{\partial z} \right |_{z=0} \simeq - \frac{B_\phi^+}{\mu_o h}
\ee
 we obtain an estimate of the toroidal field at the disc surface
\be
q \simeq \frac{1}{\alpha'_m}
\label{eq:q}
\ee  
This is an important constraint, known in the field of wind launching from thin accretion discs and worth to comment. The toroidal field at the disc surface is critical for at least two reasons. First, it is responsible for the torque due to the wind and determines therefore the disc accretion rate. Second, it conveys the power (as the MHD Poynting flux leaving the disc) that drives the wind. Its magnitude is therefore a key ingredient in magnetic jet acceleration models. The above result, due to the induction equation (\ref{eq:ind1}), tells us that the twisting of the magnetic field is stronger as the MHD turbulent diffusivity gets smaller. However, no steady-state model can be achieved with $q\gg 1$. As $\alpha'_m$ decreases, the magnetic compression of the disc becomes overwhelmingly large (Appendix \ref{AppB}) and the whole structure gets unsteady \citep{zann07}.

\bibliographystyle{/sw/share/texmf-dist/tex/latex/aa/bibtex/aa}

\end{document}